\documentclass[11pt,letterpaper]{article}
\pdfoutput=1
\usepackage{soul}
\usepackage{jheppub}
\usepackage{graphicx}
\usepackage{epstopdf}
\usepackage{float}
\usepackage{color}
\usepackage[toc,page]{appendix}
\usepackage[usenames,dvipsnames]{xcolor}

\usepackage{mathrsfs}
\usepackage[normalem]{ulem}

\newcommand{\be}{\begin{equation}}
\newcommand{\ee}{\end{equation}}
\newcommand{\ba}{\begin{eqnarray}}
\newcommand{\ea}{\end{eqnarray}}

\newcommand{\beq}{\begin{equation}}
\newcommand{\eeq}{\end{equation}}
\newcommand{\beqa}{\begin{eqnarray}}
\newcommand{\eeqa}{\end{eqnarray}}
\newcommand{\nn}{\nonumber}

\begin{document}

\title{ \boldmath On Taking the $D\to 4$ limit of Gauss--Bonnet Gravity: Theory and Solutions}

\author[a]{Robie A. Hennigar}
\author[b,c]{David Kubiz\v n\'ak}
\author[c,b]{Robert B. Mann}
\author[c]{Christopher Pollack}

\emailAdd{rhennigar@mun.ca}\emailAdd{dkubiznak@perimeterinstitute.ca}
\emailAdd{rbmann@sciborg.uwaterloo.ca}
\emailAdd{cajpollack@edu.uwaterloo.ca}

\affiliation[a]{Department of Mathematics and Statistics, Memorial University of Newfoundland, St. John's, Newfoundland and Labrador, A1C 5S7, Canada }

\affiliation[b]{Perimeter Institute, 31 Caroline St. N., Waterloo,
Ontario, N2L 2Y5, Canada}
\affiliation[c]{Department of Physics and Astronomy, University of Waterloo,
Waterloo, Ontario, Canada, N2L 3G1}

\date{April 20, 2020}

\abstract{
We comment on the recently introduced Gauss--Bonnet gravity in four dimensions.
We argue that it does not make sense to consider this theory to be defined by a set of $D\to 4$ solutions of the higher-dimensional Gauss--Bonnet gravity. We show that a well-defined
 $D\to 4$ limit of Gauss--Bonnet Gravity is obtained
generalizing a method employed  by Mann and Ross to obtain a limit of the Einstein gravity in $D=2$ dimensions.
This is a scalar-tensor theory of the Horndeski type obtained by a dimensional reduction methods.  By considering simple spacetimes beyond spherical symmetry (Taub--NUT spaces) we show that the naive limit of the higher-dimensional theory to $D=4$ is not well defined and contrast the resultant metrics with the actual solutions of the new theory. }
\maketitle

\section{Introduction}

Recently there has been lot of interest in formulating a four-dimensional Gauss--Bonnet theory of gravity.
If possible, such a theory would disqualify the privileged role of Einstein's General Relativity as a unique Lagrangian theory in four dimensions that yields
second order equations of motion for the metric.
In an intriguing recent paper \cite{Glavan:2019inb}, a new theory was
formulated by considering a $D\to 4$ limit of solutions of the $D$-dimensional Gauss--Bonnet gravity, upon rescaling the Gauss--Bonnet dimensional coupling constant $\alpha$  according to
\be\label{alpha}
(D-4)\alpha\to \alpha\,.
\ee
In this way a number of enhanced symmetry $D=4$ metrics were obtained, each carrying an imprint of higher-curvature corrections inherited from their higher-dimensional counterparts. These include
 spherical black holes \cite{Glavan:2019inb,Kumar:2020uyz,Fernandes:2020rpa,Kumar:2020owy,Kumar:2020xvu},  cosmological solutions \cite{Glavan:2019inb, Li:2020tlo, Kobayashi:2020wqy}, star-like solutions  \cite{Doneva:2020ped}, radiating solutions \cite{Ghosh:2020vpc}, collapsing solutions \cite{Malafarina:2020pvl},
with extensions to more higher-curvature Lovelock theories \cite{Casalino:2020kbt, Konoplya:2020qqh}. There are already a number of studies of the thermodynamic behaviour \cite{EslamPanah:2020hoj,Konoplya:2020cbv,Wei:2020poh,HosseiniMansoori:2020yfj,Zhang:2020qam,Singh:2020xju,Hegde:2020xlv}   and
physical properties \cite{Konoplya:2020bxa, Zhang:2020qew, Zhang:2020sjh, Wei:2020ght,Guo:2020zmf,Jin:2020emq,Heydari-Fard:2020sib,Liu:2020vkh,Islam:2020xmy, Roy:2020dyy, NaveenaKumara:2020rmi, Mishra:2020gce}
of these objects.

However, an existence of limiting solutions does not really imply an existence of a four-dimensional theory as advertised in \cite{Glavan:2019inb}, and a number of objections have been raised in this regard \cite{Gurses:2020ofy, Ai:2020peo, Shu:2020cjw},   concluding that there is no `pure four-dimensional Gauss--Bonnet gravity'.

One way \cite{Lu:2020iav} (see also \cite{Kobayashi:2020wqy}) to make a purely four-dimensional theory is via the``Kaluza--Klein-like'' procedure, namely to compactify the $D$-dimensional Gauss--Bonnet gravity on a $(D-4)$-dimensional maximally symmetric space, rescaling the coupling $\alpha$ according to \eqref{alpha}, and taking the $D\to 4$ limit, in a way reminiscent  of  how a $D\to 2$ limit of the Einstein theory was obtained many years ago \cite{Mann:1992ar}.
The resultant four-dimensional scalar-tensor theory is special in many ways. It is a special case of the Horndeski %/EdGB
theory \cite{horndeski1974second} that is obtained from a ``fundamental higher-dimensional theory'', and yields %(see also \cite{}), \rbm{What is this reference?}
four-dimensional spherical black hole solutions whose metrics coincide with the naive $D\to 4$ limit of \cite{Glavan:2019inb}.

In this paper we comment on several aspects of the new lower-dimensional Gauss--Bonnet gravity. First, we show that the same theory \eqref{SD} can be obtained as a $D\to 4$ limit of Gauss--Bonnet gravity
without reference to any choice of higher-dimensional spacetime, using an approach similar to that
employing in obtaining the $D\to 2$ limit of general relativity~\cite{Mann:1992ar}. Contrary to generic  Horndeski theory, % EdGB,
the corresponding metric is (similar to Schwarzschild) characterized by one metric function and is known analytically.
The action is
\ba\label{SD}
S&=&\int d^D x \sqrt{-g}\Bigl[R-2\Lambda+\alpha\Bigl(\phi {\cal G}+4 G^{ab}\partial_a \phi \partial_b \phi-4(\partial \phi)^2 \Box \phi+2((\partial\phi)^2)^2 \Bigr) \Bigr]\,,  \\
{\cal G}&=&R_{abcd}R^{abcd}-4 R_{ab}R^{ab}+R^2\,,
\label{GBinv}
\ea
and is applicable not only in $D=4$ but also in lower dimensions.  We believe this is the theory one should be considering when discussing the `Gauss--Bonnet gravity' in $D < 5$ dimensions.

Second, we comment on the `coincidence' of the naive $D\to 4$ limit of the Gauss--Bonnet solutions and the corresponding solutions of the theory \eqref{SD}.  Not surprisingly the results depend on the choice of higher-dimensional space; denoting by $\lambda$ the curvature of the internal space,
the additional terms
\be\label{SDlam}
S_\lambda = \int d^D x \sqrt{-g} \Bigl[-2\lambda R e^{-2\phi}-12\lambda(\partial \phi)^2e^{-2\phi}-6\lambda^2 e^{-4\phi} \Bigr]
\ee
also will appear in the action.   However for   `less-symmetric' solutions, in particular Taub-NUT spacetime as an example, we show that the theory \eqref{SD}
%(with or without \eqref{SDlam}) \tcr{\bf we are not sure about the latter I giess so let's eave the bracket out?}
does not admit the Taub-NUT solutions in the class of spacetimes with one metric function, whereas the naive $D\to 4$ limit produces such metrics.

We conclude with remarks on `uniqueness of the above limits'.  We point out that   the approach advocated in~\cite{Glavan:2019inb} of taking dimension-dependent limits of   field equations suffers from a lack of uniqueness; indeed it is not clear that it is always well defined.

%%%%%%%%%%%%%%%%%%%%%%%%%%%%%%%%%%%%%%%%%%%%%%%%%%%%%%
%%%%%%%%%%%%%%%%%%%%%%%%%%%%%%%%%%%%%%%%%%%%%%%%%%%%%%%%
\section{The $D\to 4$ limit of Gauss--Bonnet gravity}

We demonstrate here how a $D\to 4$ limit of Gauss--Bonnet gravity can be obtained without making any assumptions about the form of a higher-dimensional spacetime.   We begin by briefly reviewing the
dimensional reduction method \cite{Lu:2020iav, Kobayashi:2020wqy} to provide a contrast to our approach.

%%%%%%%%%%%%%%%%%%%%%%%%%%%%%%%%%%%%%%%%%%%%%%%%%%%%%
\subsection{Limit of vanishing compactification}

The EdGB-like action \eqref{SD} can be derived from compactifying $D$-dimensional Einstein--Gauss--Bonnet-$\Lambda$ gravity
\be\label{EGB}
S_D=\int d^D x \sqrt{-g}\bigl(R-2\Lambda+\hat \alpha {\cal G}\bigr)\,
\ee
in the limit of vanishing compactified dimensions \cite{Lu:2020iav, Kobayashi:2020wqy} (see also previous \cite{VanAcoleyen:2011mj, Charmousis:2014mia, Charmousis:2012dw, Maeda:2006hj}). Namely, the idea is to consider a Kaluza--Klein ansatz\footnote{In the above the vector part of the ansatz is switched off.  When the vector part is considered, the reduction of the Gauss--Bonnet theory yields a certain type of non-linear electromagnetism \cite{muller1988non,huang1988kaluza,soleng1995modification}, see also \cite{Gibbons:2000xe}.}
\be
ds^2_D=ds_d^2+e^{2\phi}d\Sigma_{D-d}^2\,,
\ee
where $d\Sigma_{D-d}^2$ is the line element on the internal maximally symmetric space (which we take here to be flat) and $\phi$ an additional metric function that depends only on the external $d$-dimensional coordinates.

Denoting
\be
\alpha=\epsilon \hat \alpha\,,\quad \epsilon=(D-d)\,,
\ee
the reduced $d$-dimensional action reads \cite{Lu:2020iav}
\ba\label{limitSd}
S_d^\epsilon&=&\int d^dx \sqrt{-g}e^{\epsilon\phi}\Bigl[R-2\Lambda+\epsilon(\epsilon-1)(\partial\phi)^2+\frac{\alpha}{\epsilon} {\cal G} \nonumber\\
&&\quad -\alpha (\epsilon-1)\Bigl(4G^{ab}\partial_a\phi \partial_b \phi+
2(\epsilon-2)(\partial \phi)^2\Box\phi +(\epsilon-1)(\epsilon-2)((\partial\phi)^2)^2\Bigr)\Bigr]\,.
\ea
%\ba
%S_d^\epsilon&=&\int d^dx \sqrt{-g}e^{\epsilon\phi}\Bigl[R-2\Lambda+\epsilon(\epsilon-1)(\partial\phi)^2+\frac{\alpha}{\epsilon} {\cal G} -\alpha %(\epsilon-1)\Bigl(4G^{ab}\partial_a\phi \partial_b \phi+
%2(\epsilon-2)(\partial \phi)^2\Box\phi \nonumber\\
%&&+(\epsilon-1)(\epsilon-2)((\partial\phi)^2)^2\Bigr)\Bigr]\,.
%\ea
Since the Gauss--Bonnet term
\be
\tilde S_d^\epsilon=\frac{\alpha}{\epsilon}\int d^dx \sqrt{-g}{\cal G}
\ee
is topological in $d=4$ and identically vanishes in $d<4$, one can define
\be
S_d=\lim_{\epsilon\to 0} \bigl(S_d^\epsilon-\tilde S_d^\epsilon\bigr)\,,
\ee
which yields the action \eqref{SD} after discarding total divergences.  If the metric $d\Sigma_{D-d}^2$ has constant curvature $\lambda$ then
the additional contribution \eqref{SDlam} to the action is obtained.

\subsection{The $D\to 2$ limit of General Relativity}

The method for obtaining the $D\to 4$ limit of Gauss--Bonnet gravity  without any choice of a higher-dimensional spacetime on which to compactify is similar  to that employed many years ago by Mann and Ross \cite{Mann:1992ar}, who considered a certain $D\to 2$ limit of Einstein
gravity\footnote{This approach was recently `rediscovered' \cite{Nojiri:2020tph,Ai:2020peo} in considering
generalization of the $D=4$ Gauss--Bonnet solutions.}
 that resulted in the following scalar-tensor theory:
\be\label{S2}
S^{\mbox{\tiny EH}}_2=\kappa \int d^2x \sqrt{-g}\Bigl[\psi R+\frac{1}{2}(\partial \psi)^2\Bigr]\,,
\ee
whose black hole solutions coincided with the naive $D\to 2$ limit of the $D$-dimensional solutions of the Einstein theory \cite{Mann:1989gh,Sikkema:1989ib,Mann:1991md,Mann:1991qp,Mann:1991ny,Christensen:1991dk,Lemos:1993hz,Lemos:1994fn,Mignemi:1994wg,Mureika:2011py,Frassino:2015oca}.

This theory \eqref{S2} was derived by the following conformal trick. Consider the Einstein--Hilbert action in $D$ dimensions
\be\label{EH}
S_D^{\mbox{\tiny EH}}=\hat \kappa \int d^Dx \sqrt{-g}R\,,
\ee
which is topological in $D=2$ dimensions.  Evaluating this for  a conformally rescaled metric
$\tilde g=e^{\psi} g$,
we consider the difference \cite{Mann:1992ar}
\ba
S_D^{\mbox{\tiny EH}}&=&\hat \kappa \left(\int d^Dx \sqrt{-\tilde g}\tilde R - \int d^Dx \sqrt{-g}R\right)
\nonumber\\
&=& \hat \kappa \int d^{D}x \left[e^{\epsilon \psi/2}\Bigl[\Bigl(R-(\epsilon+1)\Box \psi\Bigr)-\frac{1}{4}\epsilon(\epsilon+1)(\partial \psi)^2\Bigr] - R \right]\,,
\ea
where $\epsilon=D-2$.  Expanding the action in powers of epsilon,  and rescaling the coupling constant according to $\kappa=\epsilon \hat \kappa/2$,  we obtain
\be
S_2^{\mbox{\tiny EH}} \equiv \lim_{\epsilon\to 0} S_D^{\mbox{\tiny EH}} =
 \kappa\int d^2x \sqrt{-g}\Bigl[\psi R-\psi \Box \psi-\frac{1}{2} (\partial \psi)^2\Bigr]
=\kappa \int d^2x \sqrt{-g}\Bigl[\psi R+ \frac{1}{2} (\partial \psi)^2\Bigr]\,,
\ee
upon discarding   total divergences.

If a cosmological constant term were included in \eqref{EH}, then the additional term
\be
S_\Lambda =  -2\Lambda \int d^2x \sqrt{-g} e^\psi
\ee
would appear in \eqref{S2} upon performing the same procedure. Alternatively, one can `just add' the standard
cosmological term to \eqref{S2}.\footnote{Looking at \eqref{limitSd}, it is obvious that the theory \eqref{S2} (with or without $\Lambda$) can also be derived by using the compactification trick reviewed in the previous section, as already noted in \cite{Mann:1992ar}.}

\subsection{Gauss--Bonnet Gravity in the $D\to 4$ limit}

Let us now show that the same conformal trick yields the theory \eqref{SD}, when instead of starting with the Einstein--Hilbert action \eqref{EH}, one starts with the Gauss--Bonnet action \eqref{EGB}.
For the sake of convenience we will illustrate the method only for the Gauss--Bonnet term itself.\footnote{One could apply the conformal trick to the full action.  This would yield in addition to the theory presented below a conformally transformed Einstein-Hilbert term. However, it is possible to show that the resulting theory is equivalent via field redefinitions to the one presented here.
%{\bf I am still puzzled by this but believe you for the moment!}
}

For the purpose of applying the Mann-Ross prescription to the Gauss--Bonnet term, it is convenient to note the transformation property of the Riemann tensor under a conformal transformation $g_{ab} \to  e^{s \psi} g_{ab}$:
\begin{align}
\tilde{R}_{ab}^{cd} &= e^{-s \psi} \left[R_{ab}^{cd} + s^2 \delta_{[a}^{[c} \nabla_{b]} \psi \nabla^{d]}\psi - 2 s \delta_{[a}^{[c} \nabla^{d]} \nabla_{b]} \psi - \frac{s^2}{2} \delta_{[a}^c \delta_{b]}^d (\nabla \psi)^2  \right] \, .
\end{align}
From this it is straightforward --- if somewhat messy --- to work out the resulting conformally transformed Gauss--Bonnet density. We omit the explicit details of this process and instead move directly to discuss the limit.

Now, the term that we wish to study is topological in $D = 4$ and therefore our expansion parameter $\epsilon = (D-4)$.   We consider
\be
S_D^{GB} =  \hat \alpha \left(\int d^Dx \sqrt{-\tilde g}\tilde{\cal G} - \int d^Dx \sqrt{-g}{\cal G}\right)\,,
\ee
where $\tilde{\cal G}$ is obtained from the conformal  transformation of \eqref{GBinv}, where $g_{ab} \to e^{\psi} g_{ab}$.   Setting
$ \alpha = (D-4)\hat{\alpha} $,  we find in the $D \to 4$ limit the following theory
\begin{align}
S_4^{\cal G} &= \lim_{D\to 4} S_D^{GB} = \alpha  \int d^4 x \sqrt{-g} \bigg[ \frac{1}{2}  \psi {\cal G} - G^{ab} \nabla_a \psi \nabla_b \psi - \frac{1}{8}\left( (\nabla \psi)^2 \right)^2
-  \frac{1}{2} (\nabla \psi)^2 \Box\psi  \bigg]\,,
\label{GBD4act}
\end{align}
where we have  used the contracted Bianchi identity and the relation $\nabla^b \Box \psi - \nabla^a \nabla_b \nabla_a \psi = - \nabla_a \psi R^{a}{}_b$, along with
integration by parts, discarding  total derivatives. We note that this approach requires no assumptions about particular solutions of the original theory, nor about any background spacetime on which to compactify.

This result is remarkably similar to that in Eq.~\eqref{SD}. In fact, it is easy to show that upon setting
\be
\psi \to -2 \phi \, , \quad g_{ab} \to -\frac{1}{2} g_{ab} \, , \quad  \alpha \to \alpha/2 \, , \quad \lambda = 0 \,,
\ee
the reduced Gauss--Bonnet term presented here coincides exactly with that in Eq.~\eqref{SD}. In other words, the conformal trick has reproduced the part of the Kaluza-Klein reduced theory that does not explicitly depend on the curvature of the extra dimensions.

It is somewhat surprising that the two methods yield theories that are equivalent up to (almost trivial) field redefinitions. The approaches are rather different in spirit: In the Kaluza-Klein approach one must assume a particular form for geometry upon which the dimensional reduction is carried out. The conformal trick makes no {\it a priori} assumptions about the details of the higher-dimensional geometry. It is easy to imagine numerous alternatives to the dimensional reduction procedure that would yield likely many modifications to the theory of interest.  For example one could consider two different scalars, each respectively multiplying the two metrics
$ds_d^2$ and $d\Sigma_{D-d}^2$, yielding a different Einstein-Gauss--Bonnet dilaton gravity theory.

Modifications of the conformal trick seem less straightforward --- a natural generalization would be to consider a more general field redefinition than a conformal transformation and take a dimension-dependent limit of the redefined theory. The extent to which this approach would yield sensible modifications to the limiting theory is unclear to us, but we believe it merits further consideration.

\subsection{Equations of motion}
%\tcr{Although for simple high symmetric metrics it is possible to work directly with the effective action derived from \eqref{SD} solutions
 It is useful to present  the covariant field equations following from the action~\eqref{SD} supplemented with \eqref{SDlam}. The equation of motion for the scalar is given by
\begin{align}
{\mathcal E}_{\phi} &= - {\cal G} + 8 G^{ab} \nabla_b \nabla_a \phi + 8 R^{ab} \nabla_a \phi \nabla_b \phi - 8 (\Box \phi)^2
+ 8 (\nabla \phi)^2 \Box \phi + 16 \nabla^a \phi \nabla^b \phi \nabla_b \nabla_a \phi
\nonumber\\
&\quad + 8 \nabla_b \nabla_a \phi \nabla^b \nabla^a \phi - 24 \lambda^2 e^{-4 \phi}
- 4 \lambda R e^{-2 \phi} + 24 \lambda e^{-2 \phi} \left[ (\nabla \phi)^2 - \Box \phi \right]  \nonumber\\
&= 0\,,
\label{FEs}
\end{align}
while the variation with respect to the metric yields
\begin{align}\label{FEmet}
{\mathcal E}_{ab} &= \Lambda g_{ab} +  G_{ab} + \alpha \bigg[\phi H_{ab}  -2 R \left[(\nabla_a \phi)(\nabla_b \phi) + \nabla_b \nabla_a \phi \right] + 8 R_{(a}^c \nabla_{b)} \nabla_c \phi + 8 R_{(a}^c (\nabla_{b)}\phi) (\nabla_c \phi)
\nonumber\\
&- 2 G_{ab} \left[(\nabla \phi)^2 +  2\Box \phi \right]   - 4 \left[ (\nabla_a \phi)(\nabla_b \phi) + \nabla_b \nabla_a \phi \right] \Box \phi - \left[g_{ab}(\nabla \phi)^2 -4(\nabla_a \phi)(\nabla_b \phi) \right](\nabla \phi)^2
\nonumber\\
&+ 8 (\nabla_{(a} \phi) (\nabla_{b)} \nabla_c \phi ) \nabla^c \phi - 4 g_{ab} R^{cd} \left[\nabla_c \nabla_d \phi + (\nabla_c \phi)(\nabla_d \phi) \right] + 2 g_{ab} (\Box \phi)^2 - 2 g_{ab} (\nabla_c \nabla_d \phi)(\nabla^c \nabla^d \phi)
\nonumber\\
&- 4 g_{ab} (\nabla^c \phi ) (\nabla^d \phi) (\nabla_c \nabla_d \phi) + 4 (\nabla_c \nabla_b \phi)(\nabla^c \nabla_a \phi) + 4 R_{acbd} \left[(\nabla^c \phi)(\nabla^d \phi) + \nabla^d \nabla^c \phi \right]
\nonumber\\
&+ 3 \lambda^2 e^{-4 \phi} g_{ab}- 2 \lambda e^{-2 \phi} \left( G_{ab} + 2 (\nabla_a \phi)(\nabla_b \phi) + 2 \nabla_b \nabla_a \phi - 2 g_{ab} \Box \phi + g_{ab} (\nabla \phi)^2 \right)  \bigg] \nonumber\\
&= 0\,,
\end{align}
where in the above
\begin{align}
H_{ab} = 2 \left[R R_{ab} - 2 R_{a cd b}R^{cd} + R_{acde}R_b{}^{cde}-2R_{ac}R^c_b - \frac{1}{4} {\cal G} g_{ab} \right]\,,
\end{align}
which identically vanishes in four dimensions and less.

Interestingly, independent of $\lambda$, the above field equations satisfy
\be\label{trFeq}
0= g^{ab} \mathcal{E}_{ab} + \frac{\alpha}{2} {\mathcal E}_{\phi} = 4 \Lambda - R - \frac{\alpha}{2} \mathcal{G}  \,
\ee
in four dimensions. In other words, the trace of the field equations can be cast into a purely geometric form. This allows for important consistency checks of the solutions to the field equations as this quantity must vanish on-shell independent of the scalar field configuration. As a result, one can rather easily check whether or not solutions generated by the naive $D\to 4$ limit of solutions of the $D$-dimensional Gauss--Bonnet gravity are even possible solutions of this limiting theory  \eqref{SD}. For example, it is relatively easy to verify that the `rotating black hole metrics' constructed by the Newman--Janis trick in \cite{Wei:2020ght, Kumar:2020owy} are not solutions of the theory \eqref{SD}. This is not surprising as it is well known \cite{Hansen:2013owa} that the Newmann--Janis trick is not generally applicable in higher curvature theories. The rotating black hole solution in the four-dimensional Gauss--Bonnet gravity are thus yet to be found.

\section{Gauss--Bonnet Taub-NUTs in four dimensions}

\subsection{Spherical black holes}

Before we proceed to the Taub-NUT case we first consider the spherically symmetric ansatz
 \be
ds^2=-fdt^2+\frac{dr^2}{fh}+r^2d\Omega^2\,,
\ee
where $f=f(r)$ and $h=h(r)$ are two metric functions, $d\Omega^2=d\theta^2+\sin^2\!\theta d\varphi^2$\,, and the solution is supported by a scalar field $\phi=\phi(r)$. For simplicity, in this section we consider only the case of $\lambda=0$.

 Our goal in this section will be to check for Schwarzschild-like solutions that satisfied the condition $h=1$. It is therefore somewhat convenient to solve directly~\eqref{trFeq} for this circumstance. We find that any metric, should it exist in the full theory, must satisfy
\be\label{trSolSSS}
 \alpha f^2 -  (r^2 + 2 \alpha) f -  \frac{\Lambda r^4}{3} + C_1 r +  r^2 - C_2 = 0 \, ,
\ee
where $C_1$ and $C_2$ are arbitrary integration constants. It is straightforward to interpret these constants,
since we must recover the Schwarzschild-AdS solution for $\alpha=0$.  This implies that $C_1=-2M$ and
$C_2=0$ in this limit.

Of course having a solution to~\eqref{trFeq} does not guarantee a full solution to the theory. Therefore let us now consider the field equations in detail and determine whether there exists a scalar that supports such a solution.

Inserting this into the field equations \eqref{FEs} and \eqref{FEmet} yields 3 differential equations
for the two metric functions and the scalar field.  These equations are more easily obtained by inserting this ansatz into the action \eqref{SD} to obtain an effective Lagrangian that,  when varied w.r.t. $f, h, \phi$ yields these same 3 equations of motion.

We  seek a solution characterized by a single metric function, and so set $h=1$.  This yields
\be
(\phi'^2+\phi'')\bigl(fr^2\phi'^2-2rf\phi'+f-1\bigr)=0
\ee
from the equation of motion following from $\delta f$ .
Solving for $\phi$ we find either $\phi=\log\bigl([r-r_0]/l\bigr)$ where $l$ and $r_0$ are integration constants, or
\be\label{phiSch}
\phi^\varsigma_\pm= \varsigma \int \frac{1\pm\sqrt{f}}{\sqrt{f}r} dr\,,   \quad  \varsigma = \pm 1\,,
\ee
where the superscipt refers to the overall sign of $\phi$ and the subscript refers to the sign
in the integrand.
In what follows we focus on the latter solution which reproduces the naive $D\to 4$ solution of the Gauss--Bonnet theory.

With this,  the $\delta \phi$ equation is automatically satisfied and the remaining equation is
\be
(r^3+2\alpha r-2\alpha r f)f'+f(r^2+\alpha f-2\alpha)+\alpha-r^2+\Lambda r^4=0\,,
\ee
which follows from the $\delta h$ variation.  It has the solution\footnote{This metric was considered already in \cite{Tomozawa:2011gp,Cognola:2013fva} as a `quantum corrected metric'.} \cite{Lu:2020iav}
\be\label{LuPang}
f_\pm=1+\frac{r^2}{2\alpha}\Bigl(1\pm \sqrt{1+\frac{4}{3}\alpha \Lambda+\frac{8\alpha M}{r^3}}\Bigr)\,,
\ee
which coincides with the metric of the $D\to 4$ limit of the $D$-dimensional Gauss--Bonnet theory;
the branch $f_-$ approaches the Schwarzschild-AdS solution in the limit $\alpha\to 0$. Note that this solution is consistent with~\eqref{trSolSSS} upon setting $C_1 = - 2 M$ and $C_2 = - \alpha$.

It is straightforward to show for the asymptotically flat solution $f_- $, with $\Lambda=0$, that
 $\phi^\varsigma_- $ falls off as $1/r$.   All other solutions are non-asymptotically flat, and all sign choices for
$\phi^\varsigma_\pm$ diverge logarithmically at large $r$.

We emphasize that~\eqref{LuPang} is not the most general solution to the theory~\eqref{EGB}. Indeed, it was demonstrated in~\cite{Lu:2020iav} that there exists another solution that does not satisfy the property $h = 1$. However, it is nonetheless an interesting (and surprising) fact that there exists a solution of~\eqref{EGB} that coincides with the naive $D \to 4$ limit of solution to higher-dimensional Gauss--Bonnet gravity. It is surprising because, as nicely discussed in~\cite{Gurses:2020ofy}, one should not expect a well-defined result when taking this singular limit of higher-dimensional Gauss--Bonnet gravity. To better understand the extent to which this property is robust, in the next subsection we examine Taub-NUT solutions to the theory~\eqref{EGB}.

\subsection{Taub-NUT generalization?}

It has been known for some time \cite{Dehghani:2005zm,Dehghani:2006aa,Hendi:2008wq} that  Lovelock gravity admits also higher-dimensional generalizations of the  four-dimensional Taub-NUT space
\be\label{TNa}
ds^2=-f(dt+2n \cos\theta d\varphi)^2+\frac{dr^2}{fh}+(r^2+n^2)d\Omega^2\,,
\ee
with $h(r)=1$ and $d\Omega^2=d\theta^2+\sin^2\!\theta d\varphi^2$.  We consider here whether or not the theory~\eqref{EGB} admits Taub-NUT solutions with the same property of $h(r) = 1$ and then comment on how this relates to the naive $D \to 4$ limit of higher-dimensional Taub-NUT solutions in Gauss--Bonnet gravity.

We are once again interested in the question of whether or not a generalization of the Taub-NUT metric exists that preserves $h=1$. We start as before by considering~\eqref{trFeq} to determine the structure of the solutions, should they exist. This time we find that the metric function must satisfy
\be\label{trFeqTN}
\frac{\alpha (3n^2 - r^2) f(r)^2 }{(n^2+r^2)r} + \frac{(n^2+r^2+2\alpha) f(r) }{r} - \frac{\Lambda r (r^2 + 6 n^2)}{3} + C_2 - r - \frac{C_1}{r} \, ,
\ee
to be a potential solution of the theory. Let us now examine the full field equations to determine any such solution exists for some choice of constants.

  Inserting  the ansatz \eqref{TNa} into the action \eqref{SD}  and assuming the solution is supported by a scalar $\phi(r)$, one obtains an effective Lagrangian which, when varied w.r.t. $f, h, \psi$
yields 3 equations of motion.  Setting $h=1$, the equation of motion following from $\delta f$ yields that
\be
(\phi'^2+\phi'') \left(f(r^2+n^2)^2\phi'^2-2rf(r^2+n^2)\phi'+f(r^2-3n^2)-n^2-r^2\right) =0\,,
\ee
which has two possible solutions for $\phi$.  One is the same as the $n=0$ case: $\phi=\log\bigl([r-r_0]/l\bigr)$, where $l$ and $r_0$ are integration constants.  Following the spherical case we discard this solution and instead consider the solutions\footnote{ In fact, the logarthmic solution is inconsistent: it is not possible to find a simultaneous solution to all three independent equations of motion in this case.}
\be\label{TNphi}
\phi^\varsigma_\pm=\varsigma\int \frac{\sqrt{f(3fn^2+n^2+r^2)}\pm rf}{f(r^2+n^2)}\,,
\quad \varsigma = \pm 1\,,
\ee
which generalize \eqref{phiSch}.

Contrary to the spherical case, however, the
equation obtained by $\delta \phi$ is no longer automatically satisfied and yields a requirement that
\be
nf(r^2+4fn^2+n^2)\bigl[2rf-f'(r^2+n^2)\bigr]=0\,,
\ee
whose most general solution is
\be\label{TNf}
f=-\frac{1}{4}\frac{r^2+n^2}{l^2}\,,
\ee
where $l$ is an integration constant. Using expressions \eqref{TNphi} and \eqref{TNf},  the remaining equation of motion becomes
\ba
&&(16\Lambda l^4-12l^2-3\alpha)r^4+(32n^2\Lambda l^4+2\alpha n^2-8n^2l^2-16l^4-8\alpha l^2)r^2\nonumber\\
&&\qquad \quad \qquad+16n^4\Lambda l^4+4n^4l^2+21\alpha n^4-40n^2\alpha l^2-16n^2l^4+16\alpha l^4=0\,.
\ea
Eliminating the coefficients of the powers of $r$ requires the following two restrictions:
\be
\Lambda=\frac{3}{16}\frac{4n^2+\alpha}{n^4}\,,\quad l=\pm n\,.
\ee
 This final solution is consistent with Eq.~\eqref{trFeqTN} provided that the constants are identified as $C_1 = 0$ and $C_2 = -l^2$. Note that the solution above belongs to the Gauss--Bonnet branch of~\eqref{trFeqTN}, i.e. the one that does not admit a smooth limit to the Einsteinian theory.

 The resultant solution does not represent a Lorentzian Taub-NUT metric. It can be understood from a cosmological perspective, provided we identify $r$ as a time coordinate, setting $r\to T = n\sinh(\tau/n)$, $t\to 2n \psi$, upon which we recover
\begin{align}
ds^2_{\mbox{\tiny cos}} &=-\frac{dT^2 n^2}{T^2+n^2}+ (T^2+n^2) \bigl(d\psi+ \cos\theta d\varphi\bigr)^2+(T^2+n^2)d\Omega^2 \nonumber \\
& = -d\tau^2 + n^2 \cosh^2(\tau/n) \left(\bigl(d\psi+ \cos\theta d\varphi\bigr)^2 + d\Omega^2\right)\,,
\end{align}
which is a de Sitter type metric whose spatial sections are expanding closed 3-spheres.
%\rbm{I am not quite sure what to make of this metric -- it is not of constant curvature.}

 Alternatively, we may perform the following Wick rotation: $n\to in$ and $t\to i\tau$, to recover
\begin{align}
ds^2 &=\frac{r^2-n^2}{n^2}(d\tau+2n\cos \theta d\varphi)^2+\frac{4n^2 dr^2}{r^2-n^2}+(r^2-n^2)d\Omega^2
\nonumber\\
& = dR^2 + n^2 \sinh^2(R /n  ) \left(\bigl(d\psi+ 2 \cos\theta d\varphi\bigr)^2 + d\Omega^2\right)\,,
\end{align}
where in the second line $\tau = n\psi$ and $r=n\cosh(R/n)$.
This solution (which exists only for constrained couplings) is in fact identical to an exact Euclidean NUT-charged solution obtained in~\cite{Bueno:2018uoy} for critical Einsteinian Cubic Gravity for a particular choice of the parameter $L$ in that work --- cf. Eq.~(72) of that work. The solution generically has a conical singularity present at $r = n$ (or $R=0$) in Euclidean signature. However, in the ECG case, the solution can be analytically continued to Lorentzian signature giving rise to a wormhole geometry, albeit one that suffers from closed timelike curves if the time coordinate is identified to eliminate Misner strings. The additional constraint, $n = \pm l$, here seems to forbid the analogous analytic continuation in this case.

The conclusion of this analysis is that setting $h=1$ is too restrictive and the Taub-NUT-like solutions in this theory are more general.
%\tcr{The same remains true also for the extended theory with $\lambdaneq 0$.}
It would be interesting to better understand those solutions.

Let us now comment on  the naive $D\to 4$ limit of   Taub-NUT solutions in higher-dimensional Gauss--Bonnet gravity \eqref{EGB}.  When extending the ansatz~\eqref{TNa} to higher dimensions there are many ways to do so. The metric should have the form
\be
ds^2 = - f \left[d t+ n A_{\mathcal{B}} \right]^2 + \frac{dr^2}{f h} + (r^2+n^2) ds^2_{\mathcal{B}} \, ,
\ee
where $\mathcal{B}$ is a K{\"a}hler-Einstein base space with line element $ds^2_{\mathcal{B}}$ and the one-form $A_{\mathcal{B}}$ is a potential for the K{\"a}hler form $J = d A_{\mathcal{B}}$ on $\mathcal{B}$. In~\cite{Dehghani:2005zm,Dehghani:2006aa,Dotti:2007az,Hendi:2008wq} solutions of this form were constructed in Gauss--Bonnet and higher-order Lovelock gravities for a variety of different bases consisting of the complex projective spaces along with products of two spheres, tori, and hyperboloids. In Appendix~\ref{TNapp} we review these solutions and present the relevant equations.

Here it suffices to point out that the naive $D \to 4$ limit of these solutions is inconsistent. For this purpose we consider two possible ways that the ansatz~\eqref{TNa} can be extended to higher dimensions. The first is by considering the base to be the complex projective space $\mathbb{CP}^k$ and the second is to consider the base to be a product of two-spheres.\footnote{We have also considered extending the ansatz~\eqref{TNa} to higher dimensions by taking a base $\mathcal{B} = \mathbb{S}^2 \times \mathbb{T}^2 \times \cdots \times \mathbb{T}^2$. The limit of the higher-dimensional field equation in this case is again different and does not coincide with the solutions to~  \eqref{SD} considered above. We also mention ``compactified Gauss--Bonnet Taub-NUT metrics'' of \cite{Maeda:2006hj}, which, however do not coincide with the solutions of the present theory either.
} %\eqref{EGB} considered above.}
In the first case the limiting form of the  {higher-dimensional field equation \eqref{EGB} with $\Lambda = 0$} is
%\rbm{Which field equation are we talking about here? And to which theory?  To \eqref{SD} or to ~\eqref{EGB}?}
\be
\frac{\alpha (3n^2 - r^2) f(r)^2 }{(n^2+r^2)r} + \frac{(n^2 + r^2 + 2\alpha) f(r)}{r} - r + \frac{2n^2 + \alpha}{2 r} - C = 0 \, ,
\ee
while in the second case it is
\be
\frac{\alpha (3n^2 - r^2) f(r)^2}{r(n^2+r^2)} + \frac{(n^2 + r^2 + 2 \alpha) f(r)}{r} - r + \frac{n^2 + \alpha}{r} - C = 0  \, .
\ee
In each case $C$ is a constant of integration that is related to the ADM mass of the solution. It is clear that the two equations above are different (compare the $\alpha$ dependence in the next-to-last term) and neither of these solutions coincides with the metric obtained for the theory~\eqref{SD}.
In fact, only the latter solution approaches the spherically symmetric metric presented in the previous subsection in the limit $n\to 0$.

This serves to illustrate an important lesson. It calls into question the approach advocated in~\cite{Glavan:2019inb} of taking dimension-dependent limits of the field equations. In the two cases we illustrated above, the limiting four-dimensional metric is always identical to~\eqref{TNa}, yet the limiting equations of motion are different. This is a consequence of the fact that there are many ways by which to extend a four-dimensional geometry to higher dimenions, and in the two examples we have illustrated here the field equations have retained
non-trivial ``information'' about the higher-dimensional space from which they came, commensurate  with the discussion in~\cite{Gurses:2020ofy}. On the other hand, this makes even more surprising the fact that the theory~\eqref{SD} admits solutions coinciding with the naive limit of spherically symmetric and cosmological solutions. We believe it would be worth better understanding implications of the theory~\eqref{SD}.

 We emphasize, however, that the above argument does not prove that there does not exist dimensionally reduced theories for which these naive limits of higher-dimensional solutions are also lower-dimensional solutions. Note that for both limiting cases presented above the equations are consistent with~\eqref{trFeqTN} for different choices of the constants $C_1$ and $C_2$ and it is only through a more careful examination of the full field equations that these naive limits are excluded as solutions to the theory. Our result shows that the situation is far subtler than suggested by the spherically symmetric analysis alone and provides a concrete example of how non-uniqueness is manifest in a program of taking naive limits of the field equations.

%\tcr{\bf We should compare our solutions with \cite{Maeda:2006hj} where they consider constant dilaton, and the corresponding solutions, including Taub-NUT, see Eqs. %(3.15)! This is rather strange as I would then expect to return back to the Einstein theory}

%\tcb{\bf [Robie: We should comment on the possibility that there does exist \textit{some} dimensional reduction that would coincide with these possiblities. For example, what would reduced on a space that was a product of spheres rather than maximally symmetric spaces? There are so many options for dimensional reduction that we cannot rule out that there won't be \textit{some} theory for which the limit of the field equations is a solution, but it is not a solution to the simplest theory.]}

\section{Conclusions}

 The recent interest in generalizing   Gauss--Bonnet gravity to four spacetime dimensions stems from
 the idea of defining a theory from a set of solutions that are obtained by a certain $D\to 4$ limit of the $D$-dimensional solutions of   Gauss--Bonnet gravity. However, this procedure is neither unique nor always possible and such a theory is thus not well defined.

We have shown that a generalization of a conformal trick used to obtain the $D\to 2$ limit of general relativity
\cite{Mann:1991ny} can
be extended to obtain a $D\to 4$ limit of Gauss Bonnet gravity, given in \eqref{SD}.
  This method does not require any assumptions
about either higher-dimensional solutions or extra spacetime dimensions (and thence also the vanishing internal space),  and is purely defined in terms of 4-dimensional quantities.

 This same theory can be obtained via a compactification of the higher-dimensional Gauss--Bonnet gravity in the limit of vanishing internal dimensions, assuming the extra dimensions are conformal to a flat space.  If the spacetime has constant curvature then the resultant action is  \eqref{SD} supplemented with
 \eqref{SDlam}, which seems to be a well defined scalar--tensor theory of the Horndeski type. %\eqref{SD}.

 We have furthermore shown that, although the naive $D\to 4$ limit works for spherically symmetric solutions in four dimensions, where it coincides with the solutions of the theory \eqref{SD}, this is no longer true for more complicated solutions, such as Taub-NUT in $D=4$.  %or lower-dimensional spherically-symmetric solutions.
 This provides a further evidence that the four-dimensional Gauss--Bonnet theory cannot be simply defined by the $D\to 4$ limit of the solutions. This fact was further elaborated on in the appendix where we have shown that dependent on the character of the extra dimensions, one obtains different limits to the Taub--NUT metric in four dimensions -- the limit is thence not unique.

We have also found similar issues occur in  lower-dimensional spacetimes in the theory \eqref{SD}. We find
novel black hole solutions to the $D=3$ field equations that are not the same as those obtained by taking a naive $D\to 3$ limit \cite{Konoplya:2020ibi}; indeed we find that these latter solutions do not satisfy the field equations of
\eqref{SD} augmented with \eqref{SDlam}.  We shall report on this elsewhere.

There is one potentially problematic point that merits further commentary. As each of the examples in the text have illustrated, the scalar field configurations that support the black hole solutions exhibit logarthmic divergences asymptotically. It was argued in~\cite{Lu:2020iav} that this is not a problem because the scalar field does not appear alone in the action. However, it is clear from an examination of~\eqref{EGB} that the scalar \textit{does} appear in the action, multiplying the Gauss--Bonnet term. However, it is true that the Gauss--Bonnet term is itself a topological, or total derivative, term in four dimensions (and strictly vanishes in lower dimensions). Hence
these solutions are still well-defined, since (via an integration by parts) only derivatives of the scalar field appear in
the action. Nonetheless, we believe this issue deserves further attention to determine whether or not these scalar configurations associated with the exact solutions are perhaps excluded by some other criterion.

\section*{Acknowledgements}

This work was supported in part by the Natural Sciences and Engineering Research Council of Canada.
R.A.H.\ is supported by the Natural Sciences and Engineering Research Council of Canada
through the Banting Postdoctoral Fellowship program
D.K.\ acknowledges the Perimeter Institute for Theoretical Physics  for their support. Research at Perimeter Institute is supported in part by the Government of Canada through the Department of Innovation, Science and Economic Development Canada and by the Province of Ontario through the Ministry of Colleges and Universities.
\\

{\em Note added.} As we completed this paper, we note the appearance of  \cite{Fernandes:2020nbq}, which has an overlap with Sec. 2 of our paper.

\appendix

\section{Taub-NUT metrics in Gauss--Bonnet Gravity}
\label{TNapp}

Here we consider (Euclidean) Taub-NUT solutions to Gauss--Bonnet gravity in higher dimensions (the Lorentzian solutions studied in the main text are obtained via $\tau \to i t$ and $n \to i n$). These solutions have been considered elsewhere~\cite{Dehghani:2005zm,Dehghani:2006aa,Dotti:2007az,Hendi:2008wq}, but here we present the solutions in somewhat simpler form and review them as relevant for the main text.

The metrics we consider are of the form
\be\label{taubHigher}
ds^2 = f(r) \left[d\tau + n A_{\mathcal{B}} \right]^2 + \frac{dr^2}{f(r)} + (r^2-n^2) ds^2_{\mathcal{B}} \, ,
\ee
where $\mathcal{B}$ is a K{\"a}hler-Einstein base space with line element $ds^2_{\mathcal{B}}$. The one-form $A_{\mathcal{B}}$ is a potential for the K{\"a}hler form $J = d A_{\mathcal{B}}$ on $\mathcal{B}$.

The first base we consider is the complex projective spaces $\mathbb{CP}^k$, as this is the most natural extension of the familiar four-dimensional Taub-NUT metric to higher dimensions. The metric and K{\"a}hler potential for the complex projective spaces can be constructed recursively according to the following rules:
\begin{align}
A_{k} &= (2k+2) \sin^2 \xi_k \left(d\psi_k + \frac{1}{2k} A_{\mathbb{CP}^{k-1}} \right) \, ,
\\
ds^2_{\mathbb{CP}^k} &= (2k+2) \left[d\xi_k^2 + \sin^2\xi_k \cos^2 \xi_k \left(d\psi_k + \frac{1}{2k} A_{\mathbb{CP}^{k-1}} \right)^2 + \frac{1}{2k} \sin^2\xi_k ds^2_{\mathbb{CP}^{k-1}} \right]
\end{align}
where
\begin{align}
A_{\mathbb{CP}^1} &= 4 \sin^2 \xi_1 d\psi_1 \, ,
\\
ds^2_{\mathbb{CP}^k} &= 4 \left[d\xi_1^2 + \sin^2 \xi_1 \cos^2\xi_1  d\psi_1^2 \right] \, .
\end{align}
Here we are using the octant coordinates for $\mathbb{CP}^k$ which have ranges $0 \le \xi_i \le \pi/2$ and $0 \le \psi_i \le 2 \pi$. The normalization of the curvature used here is that $R_{ab} = g_{ab}$ for $\mathbb{CP}^k$. Of course, $\mathbb{CP}^1 \cong \mathbb{S}^2$, and thus in the case $k = 1$ we recover the familiar Taub-NUT metric, albeit written in terms of slightly non-standard coordinates.

The metric~\eqref{taubHigher} solves the field equations of Gauss--Bonnet gravity in arbitrary even dimensions. The field equation $\mathcal{E}_r^r$ can be integrated once using the integrating factor $\left[(n-r)(n+r) \right]^{k+1}/r^2$ yielding a polynomial equation that determines $f(r)$, analogous to the Wheeler polynomial that determines $f(r)$ for static black holes in Lovelock theory. Noting that $D = 2(k+1)$ the equation determining $f(r)$ becomes:
\begin{align}
0 &= \frac{-\alpha (D-2)(D-4) \left[(D-3)r^2 + 3 n^2 \right] (n^2-r^2)^{D/2 - 3} f(r)^2 }{r}
\nonumber\\
&+  \frac{(D-2) \left[(n^2-r^2) - 2(D-4) \alpha \right] (n^2-r^2)^{D/2 - 2 } f(r)}{r}
\nn\\
&+ \int \left[\frac{\left[ D (n-r)(n+r) - (D-2)(D-4)\alpha \right](n^2-r^2)^{D/2}}{(n-r)^2(n+r)^2 r^2} \right] dr + C
 \end{align}
where $C$ is a constant of integration associated with the ADM mass of the solution.

If we perform the $D \to 4$ limiting procedure advocated for spherically symmetric and cosmological solutions to this metric we obtain the following equation:
\be
-\frac{\alpha (3n^2 + r^2) f(r)^2 }{(n^2-r^2)r} + \frac{(n^2-r^2-2\alpha) f(r) }{r} + r + \frac{2n^2 - \alpha}{2 r} + C = 0 \, .
\ee
A simple inspection of this equation reveals that this limiting solution is not consistent with the Taub-NUT metrics required by the four-dimensional limit of the theory.

There are additional ways in which the Taub-NUT metric can be extended to higher dimensions --- this is through different choices for the base manifold. While the $\mathbb{CP}^k$ base is the only one that admits regular Taub-NUT and Taub-bolt solutions in Gauss--Bonnet gravity, it will be instructive to consider an alternate choice. For that, we now consider the case where the base is a product of  $(D/2-1)$ 2-spheres with
\begin{align}
A_{\mathbb{S}^2 \times \cdots \times \mathbb{S}^2} &= \sum_{i=1}^{D/2-1} \cos \theta_i d\phi_i \, ,
\\
ds^2_{\mathbb{S}^2 \times \cdots \times \mathbb{S}^2} &= \sum_{i=1}^{D/2-1} d\theta_i^2 + \sin^2 \theta_i d\phi_i^2 \, .
\end{align}
The metric is again of the form~\eqref{taubHigher}, but now the equation determining $f(r)$ is
\begin{align}
0 &= -\frac{\alpha (D-2)(D-4) \left[(D-3)r^2 + 3 n^2 \right] (n^2-r^2)^{D/2 - 3} f(r)^2 }{r}
\nonumber\\
&+  \frac{(D-2) \left[(n^2-r^2) - 2(D-4) \alpha \right] (n^2-r^2)^{D/2 - 2 } f(r)}{r}
\nn\\
&+ \int \left[\frac{(D-2) \left[(n-r)(n+r)  - (D-4)\alpha \right](n^2-r^2)^{D/2}}{(n-r)^2(n+r)^2 r^2} \right] dr + C \, .
\end{align}
This equation is almost identical to the one arising from the $\mathbb{CP}^k$ base except for the last term in the integral. This has important consequences as now the $D \to 4$ limit of the equation, which would still coincide with the familiar Taub-NUT form of the metric, is
\be
-\frac{\alpha (3n^2 + r^2) f(r)^2}{r(n^2-r^2)} + \frac{(n^2 - r^2 - 2 \alpha) f(r)}{r} + r + \frac{n^2 - \alpha}{r} + C = 0 \, .
\ee
This equation is not the same as that obtained from the $D \to 4$ limit of the $\mathbb{CP}^k$ equations. Thus, the $D\to4$ limit of the Taub-NUT solution in Gauss--Bonnet gravity is not unique. This is consistent with the results obtained in the main text.

The take home message here is that a naive $D \to 4$ limit of a solution to the field equations is still sensitive to the properties of the higher-dimensional space in which the limit was taken. As such, it is not clear that there is much meaning to such a limit in the absence of an action or field equations that exist in the four-dimensional case.

%\bibliography{references}
%\bibliographystyle{JHEP}
%\bibliographystyle{hunsrt2}

\providecommand{\href}[2]{#2}\begingroup\raggedright\endgroup

\end{document}